\documentclass[journal]{IEEEtran}

\usepackage{amssymb}
\usepackage{amsmath}
\usepackage{url}
\usepackage{array}
\usepackage{graphics}
\usepackage{epsfig}

\newtheorem{proposition}{Proposition}

\usepackage{tikz}
\usetikzlibrary{decorations.pathreplacing}

\usepackage{cite}
\usepackage[official]{eurosym}

\usepackage{siunitx} 
\usepackage{color}
\DeclareSIUnit{\Wh}{Wh}
\DeclareSIUnit{\Wp}{Wp}
\DeclareSIUnit{\pu}{p.u.}
\DeclareSIUnit{\EUR}{\mbox{\text{\euro}}}

\usepackage{multirow}
\usepackage{xcolor}

\usepackage{acronym}
\acrodef{soc}[SOC]{State-Of-Charge}
\acrodef{soe}[SOE]{State-Of-Energy}
\acrodef{ar}[AR]{autoregressive}
\acrodef{mpc}[MPC]{Model Predictive Control}
\acrodef{bess}[BESS]{Battery Energy Storage System}
\acrodef{is}[IS]{Integrated System}
\acrodef{gcp}[GCP]{Grid Coupling Point}

\acrodef{pfr}[PFR]{Primary Frequency Regulation}
\acrodef{pcr}[PCR]{Primary Control Reserve}
\acrodef{pv}[PV]{Photovoltaic system}
\acrodef{igcc}[IGCC]{International Grid Control Cooperation}
\acrodef{dap}[DAP]{day-ahead planning}
\acrodef{hap}[HAP]{hour-ahead planning}
\acrodef{tso}[TSO]{Transmission System Operator}
\acrodef{bms}[BMS]{Battery Management System}
\acrodef{isms}[ISMS]{IS Management System}
\acrodef{rod}[RoD]{Rest of the Day}
\acrodef{fh}[FH]{First Hour}
\acrodef{res}[RES]{Renewable Energy Sources}

\acrodef{rtc}[RTC]{real-time controller}
\acrodef{ulc}[ULC]{upper level controller}

\IEEEoverridecommandlockouts

\newcommand\copyrighttext{%
\footnotesize
\centering\copyright~2019 IEEE. Personal use of this material is permitted. Permission from IEEE must be obtained for all other uses, in any current or future media, including reprinting/republishing this material for advertising or promotional purposes, creating new collective works, for resale or redistribution to servers or lists, or reuse of any copyrighted component of this work in other works.\\
Published on IEEE Transactions on Sustainable Energy, Vol.11, No.3, July 2020.}
\newcommand\copyrightnotice{%
\begin{tikzpicture}[remember picture,overlay]
\node[anchor=south,yshift=-2pt] at (current page.south) {\setlength{\fboxrule}{0pt}\fbox{\parbox{\dimexpr\textwidth-\fboxsep-\fboxrule\relax}{\copyrighttext}}};
\end{tikzpicture}%
}

\begin{document}

\title{Day-Ahead and Intra-Day Planning of Integrated BESS-PV Systems providing Frequency Regulation}

\author{Francesco~Conte,~\IEEEmembership{Member,~IEEE,}
        Stefano~Massucco,~\IEEEmembership{Senior~Member,~IEEE,}
        Giacomo-Piero~Schiapparelli,~\IEEEmembership{Student~Member,~IEEE,}
        and~Federico~Silvestro,~\IEEEmembership{Senior~Member,~IEEE}
\thanks{F. Conte, S. Massucco, G.-P. Schiapparelli, F. Silvestro are with are with the Dipartimento di Ingegneria Navale, Elettrica, Elettronica e delle Telecomunicazioni, %
Universit\`{a} degli Studi di Genova, via all'Opera Pia, 11A , I-16145 Genova (GE), Italy, e-mail: fr.conte@unige.it, stefano.massucco@unige.it, giacomo-piero.schiapparelli@edu.unige.it, federico.silvestro@unige.it \textit{Corresponding author:} Federico Silvestro}
\thanks{ \textit{Corresponding author:} Federico Silvestro}
\thanks{ DOI: 10.1109/TSTE.2019.2941369}}

\markboth{Preprint submitted to IEEE TRANS. ON SUSTAINABLE ENERGY, published June 2019}%
{Shell \MakeLowercase{\textit{et al.}}: Bare Demo of IEEEtran.cls for IEEE Journals}

\IEEEaftertitletext{\copyrightnotice\vspace{-1\baselineskip}}
\maketitle

\begin{abstract}
The paper proposes an optimal management strategy for a system composed by a battery and a photovoltaic power plant. This integrated system is called to deliver the photovoltaic power and to simultaneously provide droop-based primary frequency regulation to the main grid. The battery state-of-energy is controlled by power offset signals, which are determined using photovoltaic energy generation forecasts and predictions of the energy required to operate frequency regulation.
A two level control architecture is developed. A day-ahead planning algorithm schedules the energy profile which is traded at the day-ahead market and defines the primary control reserve that the integrated system is able to provide in the considered day. During the day operations, a second level algorithm corrects the dispatched plan using updated information, in order to guarantee a continuous and reliable service. Both control algorithms take into account the uncertainties of the photovoltaic generation and of the frequency dynamics using stochastic optimization.
\end{abstract}

\begin{IEEEkeywords}
Battery energy storage systems, primary frequency regulation, primary control reserve, predictive control, photovoltaic systems.
\end{IEEEkeywords}

\IEEEpeerreviewmaketitle

\section{Introduction}

\IEEEPARstart{T}{he} instantaneous balance between generated and consumed active power is one of the basic principles of the AC power systems operation. Any variation from such a condition causes a frequency event, namely, the deviation of the system frequency from its nominal value. 
The progressive displacement of conventional generation in favour of production from \ac{res} will cause the reduction of the frequency control capability of power systems. Therefore, it is necessary to involve new resources in grid ancillary services in order to ensure robustness, resiliency and efficiency of future power systems \cite{Ye2016,Vrettos2016,Baccino2014}.

The power equilibrium in real-time can be controlled only if the production system is able to change its generation level \cite{ucte2009appendix}. The coupling of \ac{res} with \acp{bess} is therefore investigated in order to meet the grid flexibility requirements with the aleatory characteristics of such generation systems \cite{guo2013electricity,wang2016energy,nair2010battery}. 
Assessments on the capital costs of batteries have shown that, with the market condition of last years, a multifunctional storage deployment is necessary to overcome the investment costs for energy storage systems \cite{wasowicz2012evaluating}. 

Many literature papers propose methods for allowing batteries to provide services such as energy management, peak shaving, and frequency and voltage regulation \cite{namor2018control, oudalov2007sizing, christakou2014primary, oudalov2007optimizing, BULLICHMASSAGUE201749, conte2017stochastic, Silvestro2018, Yang2018, Li2013, Lawder2014, Zamani2015, Park2017, stai2018,MOHAGHEGHI2018,MOHAGHEGHI2018a}. 
Several control strategies to perform \ac{pfr} are proposed in literature \cite{lu2014state, megel2013maximizing, khalid2010model}. Moreover, specific markets around the world are now under development in order to integrate \ac{bess} into grid services, such as in the United States PJM interconnect and ISO New England \cite{PJMm12_2019,PJMm18_2019}, in the Europe National Grid (GB) \cite{EFR_Uk} and in the \ac{igcc} which involves German, Belgian, Dutch, French, Swiss and Austrian \ac{pcr} markets \cite{internationalPCR}.

In this work an integrated \ac{bess}-\ac{pv} system is considered.
A wide literature shows how to properly manage this \ac{is} to perform multiple services such as contingency management, peak shaving, demand response, etc. \cite{Shi2018,Perez2016,eyer2010energy}. However, in many cases droop-based \ac{pfr} is not considered.
Papers combining multiple services with \ac{pfr} usually assume a non-traditional provision of \ac{pfr}, such as the one defined by the PJM market \cite{PJMm12_2019}. In this specific case, the signal provided to the regulating units is divided in two contributions, a slow one (RegA) and a fast one (RegD). The one provided to BESS and \ac{res} is RegD, which is designed to be zero-mean, in order to keep the BESS \ac{soc} approximately at the same level, during a given time period \cite{Shi2018,Cheng2018,Perez2016}. Nevertheless, most markets do not adopt this control strategy, but use the row frequency as regulating signal, which is not guaranteed to be zero-mean within a given time period. In this case, more sophisticated techniques, such as the ones in \cite{PSCC2018} and \cite{namor2018control} should be used.

In particular, in \cite{PSCC2018} and \cite{namor2018control} \ac{pfr} is coupled with the dispatch of the active power demand of a distribution feeder. Moreover, such as other works previously cited, these two works are focused on the usage of batteries in transmission and distribution level. Differently, the present paper is focused on the generation level: the \ac{is} is operated as a power plant which simultaneously participates to the energy market, delivering to the grid the available \ac{pv} generation, and provide droop-based \ac{pfr}. The main contribution of this work is therefore the integration of these two services with a common formulation. Moreover, the problem is defined in order to match the current grid codes and markets requirements (see Section \ref{ssec:PFR_req} for details).

\begin{figure}[t]
	\centering
	\includegraphics[width=\columnwidth]{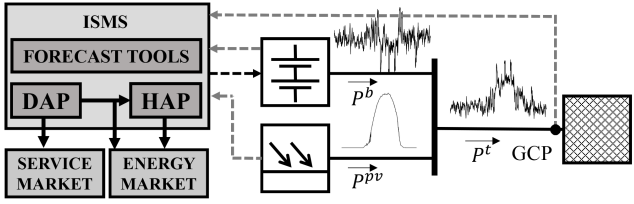}
	\caption{Integrated system configuration scheme.}
	\label{fig:Onelinediagram}
\end{figure}

The \ac{is} architecture is depicted in Fig. \ref{fig:Onelinediagram}. The objective is to define an energy dispatch plan using the storage flexibility, to maximize the economic gain and provide a continuous and reliable \ac{pfr} service.
A two level strategy \cite{Borghetti2007} is adopted. A suitably developed algorithm, called \ac{dap}, defines an energy dispatch plan and a droop coefficient for the up-coming day, both traded at the day-ahead market. \ac{dap} uses the forecasts of the \ac{pv} generation and of the energy required to perform \ac{pfr}. The latter information is provided by a method proposed in \cite{PSCC2018}. Then, during the day operation, an \ac{hap} algorithm corrects the \ac{dap} dispatch plan using updated short-term forecasts and the current battery \ac{soe}, in order to assure the continuity of the \ac{pfr} service. The dispatch plan corrections are traded at the intra-day energy market. Both \ac{dap} and \ac{hap} use chance-constrained optimization \cite{Cinquemani2011}, in order to take into account the uncertainties of the \ac{pv} generation and of the frequency signal dynamics.

It is worth remarking that the problem formulation is general, there are no hypotheses on the type of battery or its performance or the ratings of the resources. Moreover, there are neither hypothesis on the coupling between the \ac{bess} and the \ac{pv} plants, that could be in principle in AC, DC or even the results of an aggregation of several \acp{bess} and \acp{pv}.

The performances of the designed method are tested by simulations in MATLAB/Simulink, the test environment adopted has been validated by on field experiments as detailed in \cite{PSCC2018}.

The rest of the paper is organized as follows.  Section \ref{sec:ProblemDefinitions} describes the system configuration and provides the problem formulation. Section~\ref{sec:DAP} and Section~\ref{sec:HAP} introduced the \ac{dap} and \ac{hap} algorithms, respectively. Simulation results are described in Section~\ref{sec:simulations}.  Finally, conclusions are reported in Section~\ref{sec:conclusions}.

\vspace{10pt}
\textit{Notation.} $\mathbf{E}(z)$ is the expectation of the random variable $z$; $\mathbf{P}(A)$ is the probability of event $A$; $x \sim\mathcal{N}(\bar{z},\sigma^2)$ indicates that $z$ is a Normally distributed random variable with mean 
$\bar{z}$ and variance $\sigma^2$; ${\rm erf}^{-1}(\cdot)$ is the {inverse Gauss error function}; $k=a:b$, denotes the sequence $k=a,a+1,\ldots,b$.    

\section{Problem Formulation} \label{sec:ProblemDefinitions}
The system configuration is presented in Fig.~\ref{fig:Onelinediagram}. The \ac{is} is composed by a \ac{bess} and a \ac{pv} plant. The power $P^t$ [kW] is exported at the \ac{gcp}. As indicated, $P^t>0$ means that the \ac{is} is exporting power. With the same convention, the \ac{bess} exports or import power $P^b$ [kW] and the \ac{pv} plant generates power $P^{pv}$ [kW]. From the figure, it clearly follows that
\begin{equation}
    P^t = P^b + P^{pv}.
\end{equation}

The \ac{pv} generation and the \ac{bess} power exchange are limited by the rated powers $P^{pv}_{\rm n}$ and $P^{b}_{\rm n}$, respectively. The \ac{is} rated power is indicated with $P^{t}_{\rm n}=P^{pv}_{\rm n} + P^{b}_{\rm n}$. The \ac{bess} energy capacity is indicated with $E_{\rm n}$ [kWh].  

The \ac{is} has the objective of exporting the \ac{pv} generation and provide \ac{pfr}. Therefore, $P^t$ assumes the form
\begin{equation}\label{eq:Pg}
P^t = P^m - \alpha \Delta f ,
\end{equation}
where $\alpha$ [kW/Hz] is the \textit{droop} coefficient, $\Delta f$ [Hz] is the frequency deviation from the nominal value $f_{\rm n}$ and $P^m$ [kW] is the \ac{is} \textit{market} power, \textit{i.e.} the power traded at the energy market. The duration of the energy market sessions, also called dipatch sampling time, will be indicated with $\tau$ [\si{\second}]. 

It is assumed that the \ac{is} always operates as a generator, and therefore $P^m\geq0$. A minimal droop coefficient $\alpha_{\min}$ is established. It is therefore required that
\begin{equation}\label{eq:alphamin}
\alpha \geq \alpha_{\min}.
\end{equation}

The value of $\alpha_{\min}$, can be defined, for example, according to \cite{aisbl2012entso}, where a generator with rated power $P_{\rm n}$ participating to \ac{pfr} has to ensure a maximum statism $b_p^{\max}$ [\si{\percent}], that corresponds to $\alpha_{\min}$ by the relation
\begin{equation}\label{eq:statism_max}
b_p^{\max} = \frac{100}{\alpha_{\min}}\cdot \frac{P_{\rm n}}{f_n}.
\end{equation}

\ac{pfr} is effectively operated only by \ac{bess}. Therefore, to obtain \eqref{eq:Pg}, it results that the battery power exchange is
\begin{equation}\label{eq:Pb}
P^b = P^m - P^{pv}  - \alpha \Delta f.  
\end{equation}

The \ac{is} is controlled by a \ac{isms} that receives measurements and sends control set-points from/to the \ac{pv} inverter and the \ac{bms}, which controls the \ac{bess}. In particular, the \ac{isms} receives the measurements of the current \ac{pv} power generation $P^{pv}$ and of the battery State-of-Energy, indicated with $S$ [p.u.].

In this paper, the \ac{soe} dynamics is modelled by the following discrete-time system:
\begin{align}
    \label{eq:SEOdyn}
&S_{k+1} = S_k - \frac{\tau}{3600 \cdot E_{\rm n}} P^b_k. 
\end{align}
Notice that \eqref{eq:SEOdyn} describes the dynamics of a \ac{bess} with unitary efficiency. It will be shown that such an assumption in the control algorithm design do not affect the overall results. The same approximation has been done and verified in \cite{PSCC2018,powertech2019}

The \ac{isms} has the mission of maximizing the economic gain coming from the energy delivery and the provision of the \ac{pfr} service. It uses forecasts of the \ac{pv} generation and of the energy required to provide \ac{pfr}. Based on this information, each day, the \ac{isms} trades the energy delivery profile and the day \ac{pfr} droop coefficient $\alpha$ for the day-ahead. During the operation the battery \ac{soe} must be kept within the \textit{security interval} $[S^{\rm min},S^{\rm max}]$. The violation of the \ac{soe} security interval is called \textit{failure}. When a failure occurs, the provision of \ac{pfr} is suspended. The percentage time during which the \ac{soe} security interval is violated is defined \textit{failure rate}, indicated with $\lambda$.

Using stochastic modelling, the \textit{a priori} definition of maximal failure rate is, for all $k$, 
\begin{equation}
    \lambda_{\max} = 1-\mathbf{P}\left( S^{\min} \leq S_k \leq S^{\max} \right),
\end{equation}
\textit{i.e.} the probability of violation of the security interval. 

The \ac{dap} is operated by a properly developed optimization algorithm which has the objective of maximizing the economic gain and simultaneously assuring that $\lambda$ is lower than a predetermined maximal value $\lambda_{\rm max}$. The \ac{dap} program can be applied directly; however, a second possibility is proposed. Indeed, during the day, using updated short-term forecasts, it is possible to operate corrections to reduce the failure rate. This is realized by the \ac{hap} algorithm.

Both \ac{dap} and \ac{hap} algorithms use the technique introduced in \cite{PSCC2018} for providing \ac{pfr} from BESSs. Therefore, before introducing \ac{dap} and \ac{hap}, the technique proposed in \cite{PSCC2018} is briefly recalled in the following.

\subsection{Primary frequency regulation from \ac{bess}}\label{sec:FrequencyForecast}
Assume to have a \ac{bess} with capacity $\tilde{E}_{\rm n}$ which performs \ac{pfr} with a droop coefficient $\alpha$, and divide the time into windows of length $T$ [\si{\hour}]. The energy required to provide \ac{pfr} in the generic $i$-th time window $[iT,(i+1)T]$ is:
\begin{equation} \label{eq:EiPFR}
   E_i^{f} =-\alpha \cdot \int_{iT}^{(i+1)T}\Delta f(t) dt = -\alpha W^f_{i},    
\end{equation}
where $W^f_i$ [\si{\hertz\hour}] is defined as the integral over the current time interval of the frequency deviation. The analysis detailed in \cite{PSCC2018} demonstrates that a time series $\{W^f_i\}$ obtained from a large database of frequency measurements \cite{DatiFreq} and a given value of $T$ (e.g. $T \in[1, 2, \dots ,24]$\si{\hour}) can be modeled with an \ac{ar} process of order $p$ \cite{ref:MadsenTimeSeriesAnaysis}. This implies that: 
\begin{equation} \label{eq:ARmodel}
    W^f_{i+1} = \widehat{W}^f_{i+1} + \epsilon_{i+1},
\end{equation}
\begin{equation}
    \widehat{W}^f_{i+1} = W^f_{i} \phi_1 + \cdots + W^f_{i-p-1} \phi_p,
\end{equation}
where $\{W^f_{i},\dots,W^f_{i-p-1}\}$ are the measured value of the integral of the frequency deviation in the last $p$ periods,  $\{\phi_1,\dots,\phi_p\}$ are the \ac{ar} coefficients defined by the analysis of the frequency database, $\widehat{W}_i^f$ is the prediction $W^f$ for the upcoming period, and $ \epsilon_i$ is a zero-mean Gaussian random variable with standard deviation $\sigma^w_T$. The dependence on $T$ of this standard deviation is explicitly indicated with the subscript because, in the following, different values of $T$ will be used. It is worth remarking that $\sigma^w_T$ increases with $T$.

Based on this model, the following \textit{energy offset} is defined:
\begin{equation}\label{eq:EnergyOffset}
\widehat{E}^o_i = \left( S_i -\frac{1}{2} +\frac{\alpha \widehat{W}^f_i}{\tilde{E}_{\rm n}} \right)\tilde{E}_{\rm n},
\end{equation}
where $S_i$ is the battery \ac{soe} at the beginning if the $i$-th time window.
In \cite{PSCC2018} is proved that, if $\widehat{E}^o_i$ is exchanged by the \ac{bess} during $i$-th time window, then the \ac{bess} can provide \ac{pfr} with a maximal failure rate $\lambda^f_{\rm max}$, with respect the \ac{soe} the security interval $[0,1]$, if the droop coefficient $\alpha$ is equal or lower than the maximal value
\begin{equation}\label{eq:alphamax}
    \alpha_{\max} = \frac{\tilde{E}_{\rm n}}{2\cdot \mu \cdot \sigma^w_T},
\end{equation}
where $\mu$ is $(1-\lambda^f_{\max}/2)$-th percentile of a zero-mean standard Gaussian random variable, which can be computed as $\mu~=~\sqrt{2}{\rm erf}^{-1}(1-\lambda^f_{\rm max})$.

\subsection{Main requirements for PFR service}\label{ssec:PFR_req}
The integration of \ac{res} into grid regulating scheme requires the revision of the grid codes. In continental Europe, all the \acp{tso} involved in the joint market \ac{igcc} have worked together to define pre-qualification and delivery rules for the \acp{bess} which provide \ac{pcr} \cite{internationalPCR}.
In the UK, Nationalgrid (NGET) has developed the enhanced frequency response service and defined specific rules for the integration of the new resources into the markets \cite{FRserviceUSandEU}.
In the United States of America, PJM has created another market in which the users are remunerated for the capacity, for the availability and for the performance in providing the service \cite{EFR_Uk,PJMm18_2019}.

By analyzing the mentioned documents, it results that the \ac{pfr} markets are different each others and still changing, mainly because they are new. Therefore, the control strategy designed in this paper has the objective of matching the most important rules common between those markets rules: 
\begin{enumerate}
    \item[a)] droop-based response to the frequency variations;
    \item[b)] the \ac{soe} must be kept within predefined limits; 
    \item[c)] as requested by the market operators \cite{internationalPCR,FRserviceUSandEU,aisbl2012entso}, a minimum \ac{pcr} offer has to be ensured;
    \item[d)] according to some grid operators, the failure rate has to be kept lower than a maximal value (\textit{e.g.} 5$\%$ in UK \cite{DatiFreq,EFR_Uk}) or equal to zero \cite{swissgrid2017,zeh2016fundamentals,internationalPCR,PJMm18_2019} in order not to pay penalties.  
\end{enumerate}

Finally note that the algorithm proposed in the present paper does not respect the capacity trading time line, \textit{i.e.} the droop coefficient is computed daily and not weekly as in \cite{internationalPCR}. However, it is opinion of the authors that future markets deregulation will require to operate on shorter time windows in order to integrate all the new resources.

\section{Day-Ahead Planning (DAP)}\label{sec:DAP}
The \ac{dap} problem consists in the definition of the daily power
delivery profile $\{P^m_k\}$ of the \ac{is} and the droop coefficient $\alpha$, computed one day before. 
The objective is to maximize the economic gain, given set of available data and satisfying a set of technical constraints, as detailed in the following.

\subsection{Available data}
Given the time horizon $N = 24 \cdot 3600/\tau$ , the data
supposed to be available at day $d-1$ when the planning of day $d$ is computed are:
\begin{itemize}
\item[a)] a \ac{pv} forecast profile $\{\widehat{P}^{pv}_k\}_{k=0}^{N-1}$, with an associated confidence interval $\Delta^{pv}_k$, such that
$
| P^{pv}_k - \widehat{P}^{pv}_k | \leq \Delta^{pv}_k;
$
\item[b)] the prediction of the frequency integral for the day-ahead $\widehat{W}^f_{d}$ and the associated standard deviation $\sigma^w_{24}$, computed as described in Section~\ref{sec:FrequencyForecast} with $T=$\si{24}{h};
\item[c)] the energy price profile $\{c^{e}_k\}_{k=0}^{N-1}$;
\item[d)] the \ac{pfr} price $c^{f}$;
\item[e)] the day initial \ac{soe}, $S_0$.
\end{itemize}

\subsection{\ac{soe} constraints}
Based on the \ac{pv} forecast data, the \ac{pv} power profile is represented with the following Gaussian model:
\begin{equation}\label{eq:Pvgauss}
P_k^{pv} \sim \mathcal{N}(\widehat{P}_k^{pv},(\sigma^{pv}_k)^2), \ \ \sigma^{pv}_k=\Delta^{pv}_k/3
\end{equation}
so that $\mathbf{P}(\vert \widehat{P}_k^{pv}- {P}_k^{pv} \vert)\leq 0.997.$ From \eqref{eq:Pb}, \eqref{eq:SEOdyn} and definition \eqref{eq:EiPFR} (with $T=\tau$) it follows that, for $k=0:N-1$,
\begin{equation}\label{eq:SEOdyn2}
    S_{k+1} = S_k - \frac{\tau (P^m_k-P^{pv}_k) }{3600 \cdot E_{\rm n}} + \frac{\alpha  W^f_k }{E_{\rm n}}.
\end{equation}

\begin{figure}[t]
	\centering
	\includegraphics[width=0.9\columnwidth]{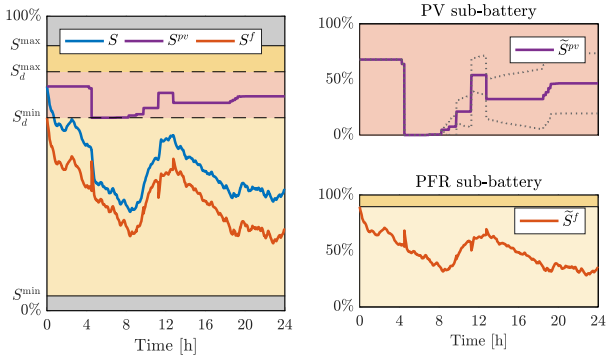}
	\caption{\ac{dap} optimization principle scheme.}\label{fig:DAPscheme}
\end{figure}

Figure \ref{fig:DAPscheme} shows the basic principle of the \ac{dap} optimization. Firstly, the equivalent \ac{bess} capacity ${E}^s_{\rm n}$ is defined as 
\begin{equation}\label{eq:Es}
    {E}^s_{\rm n} = E_{\rm n}(S^{\max}-S^{\min}).
\end{equation}
Then, each day, the quantities $S_d^{\rm max}$ and $S_d^{\min}$
are determined by the optimization, to divide ${E}^s_{\rm n}$ in two portions ${E}_{\rm n}^{pv}$ and ${E}_{\rm n}^f$:
\begin{equation} \label{eq:Epvn}
{E}_{\rm n}^{pv} = E_{\rm n}(S_d^{\rm max}-S_d^{\rm min}),
\end{equation} 
\begin{equation}\label{eq:Efn}
E_{\rm n}^{f} = E^s_{\rm n}-E_{\rm n}^{pv}.
\end{equation} 
It is obviously required that 
\begin{align}
S^{\min} \leq S_d^{\min} \leq S_d^{\max}\leq S^{\max}. \label{eq:cSminmax}
\end{align}

The idea is to use the portion $E^{pv}_{\rm n}$ to correct the \ac{pv} prediction errors, and the portion $E^f_{\rm n}$ to provide \ac{pfr}, as they were two different batteries: the \textit{PV battery} and the \textit{PFR battery}, respectively. Two equivalent \ac{soe} trajectories $\{\widetilde{S}^{pv}_k\}$ and $\{\widetilde{S}^f_k\}$ are supposed to move in these two batteries. They are defined in \textit{p.u.} with respect to the two capacities $E^{pv}_{\rm n}$ and $E^{f}_{\rm n}$ (right plots in Fig.~\ref{fig:DAPscheme}), by the following dynamical equations (with $k=0:N-1$):
\begin{align}\label{eq:SOEpv}
    &\widetilde{S}_{k+1}^{pv} = \widetilde{S}_{k}^{pv} - \frac{\tau  (P^m_k-P^{pv}_k) }{3600 \cdot E^{pv}_{\rm n}} \\
    &\widetilde{S}_0^{pv} = \frac{E_{\rm n}(S_0-S^{\rm min}_d)}{E^{pv}_{\rm n}}, \label{eq:SOEpv0} \\
    &\widetilde{S}_{k+1}^{f} = \widetilde{S}_{k}^{f} + \frac{\alpha W^f_k }{E_{\rm n}^f}, \label{eq:SOEfdyn} \\
    &\widetilde{S_0}^{f} = \frac{E_{\rm n}(S^{\rm min}_d-S^{\rm min})}{E^f_{\rm n}}, \label{eq:SOEf0}
\end{align}
It can be proved by induction that, for $k=0:N$, 
\begin{equation}\label{eq:SOEdec}
S_k =S_k^{pv}+(S_k^{f}-S^{\min}_d),
\end{equation}
where $S_k^f$ and $S_k^{pv}$ are defined as it follows (see the left plots in Fig.~\ref{fig:DAPscheme} for an example):
\begin{equation}\label{eq:SOEdec1}
S_k^{pv} =\frac{E_{\rm n}^{pv}\widetilde{S}_k^{pv}}{E_{\rm n}}+S^{\min}_d, \quad S_k^f = \frac{E^f_{\rm n}\widetilde{S}_k^f}{E_{\rm n}} + S^{\rm min}.
\end{equation}

The component $S^{pv}$ is driven by the dispatch power $P^m$ and the \ac{pv} power $P^{pv}$, whereas the component ${S}^{f}$ is driven by the frequency variations. Since the (local) \ac{pv} production and grid frequency can be assumed to be statistically independent, also ${S}^{pv}$ and ${S}^{f}$ result to be independent. This implies the following result, which is proved in the appendix section.

\begin{proposition} \label{pr:1}
If, for all $k=0:N$,
\begin{align}
&\mathbf{P}(0\leq \widetilde{S}^{pv}_k\leq 1) \geq 1-\beta,\label{eq:chanceSOEpv} \\
&\mathbf{P}(0\leq \widetilde{S}^{f}_k\leq 1) = 1-\lambda^f_{\max},\label{eq:chanceSOEf}
\end{align}
then 
\begin{equation}\label{eq:failure1}
\mathbf{P}(S^{\min}\leq S_k\leq S^{\max}) \geq 1-{\lambda}_{\max}
\end{equation}
with 
\begin{equation}\label{eq:failure2}
{\lambda}_{\rm max} = \lambda^f_{\rm max} + \beta - \lambda^f_{\rm max}\beta.    
\end{equation}
\end{proposition}

This proposition means that if \eqref{eq:chanceSOEpv} and \eqref{eq:chanceSOEf} hold true, than ${\lambda}_{\rm max}$ is the resulting maximal failure rate of the \ac{is}. 

Relation \eqref{eq:chanceSOEpv} is considered as a chance constraint.
Using the Gaussian representation \eqref{eq:Pvgauss}, assuming that the \ac{pv} prediction errors and the battery modelling errors are independent, and that the sampling time $\tau$ is large enough to suppose that the \ac{pv} prediction errors at different time steps are mutually independent, from \eqref{eq:SOEpv}--\eqref{eq:SOEpv0}, it follows that, for $k=0:N$,
\begin{equation}\label{eq:SOEpvGauss1}
    \widetilde{S}_{k}^{pv} \sim \mathcal{N}\left( m^{s}_k,(\sigma_k^s)^2 \right),
\end{equation}
where
\begin{equation}\label{eq:SOEpvGauss2}
    m^s_k = \widetilde{S}^{pv}_0 - \frac{\tau  }{3600 \cdot E_{\rm n}^{pv}} \sum_{j=0}^{k-1} (P^m_j-\widehat{P}^{pv}_j),
\end{equation}
\begin{equation}\label{eq:SOEpvGauss3}
    (\sigma_k^s)^2 = \left(\frac{\tau }{3600 \cdot E_{\rm n}^{pv}} \right)^2 \cdot \sum_{j=0}^{k-1} (\sigma_j^{pv})^2 
\end{equation}

To obtain \eqref{eq:chanceSOEpv}, the following separated chance constraints are defined, for all $k=0:N$:
\begin{align}
&\mathbf{P}\left(\widetilde{S}^{pv}_k \leq 1\right) \geq 1-\frac{\beta}{2}, \ \ \mathbf{P}\left(\widetilde{S}^{pv}_k \geq 0\right)  \geq 1 -\frac{\beta}{2} \label{eq:cc2}  
\end{align}
which, using the Gaussian model \eqref{eq:SOEpvGauss1}--\eqref{eq:SOEpvGauss3}, can be expressed with the equivalent deterministic constraints (see \cite{Cinquemani2011} or \cite{conte2017stochastic} for details):
\begin{align}
m^s_k + \theta_s \sigma^{s}_k \leq 1, \label{eq:msconstr1}\\
-m^s_k +\theta_s \sigma^{s}_k \leq 0, \label{eq:msconstr2}
\end{align}
where $\theta_s = \sqrt{2}{\rm erf}^{-1}(1-\beta)$.

To obtain \eqref{eq:chanceSOEf}, the method recalled in Section~\ref{sec:FrequencyForecast}
is applied to the \ac{pfr} battery consideiring a period $T=$\SI{24}{\hour}. Recall that $S^{\min}_d$ and $S^{\max}_d$ are defined by the \ac{dap} optimization. Considering \eqref{eq:SOEf0}, this implies that the initial condition $\widetilde{S}^{f}_0$, at the beginning of the day, is defined by the optimization. Therefore, by \eqref{eq:EnergyOffset}, if
\begin{equation}\label{eq:csoef}
    \widetilde{S}^f_0 = \frac{1}{2}-\frac{\alpha\widehat{W}^f_d}{E^f_{\rm n}},
\end{equation}
then the required energy offset $\widehat{E}^o_d=0$, and therefore \eqref{eq:chanceSOEf} is satisfied with $\alpha$ given by 
\begin{equation}\label{eq:calphad}
\alpha = \frac{E^f_{\rm n}}{2\mu\sigma^w_{24}}. 
\end{equation}

Using the definition of $E^f_{\rm n}$ in \eqref{eq:Efn} and the relation \eqref{eq:SOEf0}, it can be shown that \eqref{eq:csoef} and \eqref{eq:calphad} are equivalent to
\begin{equation}\label{eq:csoef1}
    2\alpha\widehat{W}_d^f = E_n[(S^{\rm max}+S^{\rm min}) -(S^{\rm max}_d+S^{\rm min}_d)]
\end{equation}
\begin{equation}\label{eq:calphad1}
    2\alpha \mu \sigma_{24}^w = E_{\rm n} [(S^{\max}-S^{\min})-(S^{\max}_d-S^{\min}_d)].
\end{equation}

\subsection{Power constraints}
As defined in Section \ref{sec:ProblemDefinitions}, the \ac{bess} power is limited by the nominal value $P^b_{\rm n}$. From \eqref{eq:Pb}, it results that the following inequality should be always satisfied:
\begin{equation}\label{eq:cpbmax}
    \vert P^b \vert = \vert P^m-P^{pv}-\alpha \Delta f \vert \leq P^b_{\rm n}.
\end{equation}
Since it is assumed that, for $k=0:N-1$,
\begin{equation}\label{eq:positivePm}
    0\leq P^m_k \leq P^t_{\rm n}
\end{equation}
and $P^{pv}_k\geq 0$ by definition, then, for the day-ahead $d$, there are two worst cases, which are covered with the following chance constraints (with $k=0:N-1$):
\begin{align}
    &\mathbf{P}(P^m_k-P^{pv}_k+\alpha\Delta f^{\rm max} \leq P^b_{\rm n})\geq 1-\gamma, \label{eq:p_constraint1}\\
    &\mathbf{P}(P^m_k-P^{pv}_k-\alpha\Delta f^{\rm max} \geq -P^b_{\rm n})\geq 1-\gamma. \label{eq:p_constraint2}
\end{align}
where $\Delta f^{\rm max}$ is the maximal frequency variation \cite{ucte2009appendix}. Based on the Gaussian model of the \ac{pv} forecasts \eqref{eq:Pvgauss}, \eqref{eq:p_constraint1} and \eqref{eq:p_constraint2} can be expressed with the equivalent deterministic constraints (see \cite{Cinquemani2011} or \cite{conte2017stochastic} for details):
\begin{align}
&P^m_k-\hat{P}^{pv}_k  + \alpha \Delta f_{\rm max}+\theta_b \sigma^{pv}_k\leq P^b_{\rm n}, \label{eq:p_constraint1a} \\
&P^m_k-\hat{P}^{pv}_k  - \alpha \Delta f_{\rm max}-\theta_b \sigma^{pv}_k\geq -P^b_{\rm n}, \label{eq:p_constraint2a}
\end{align}
with $k=0:N-1$, and $\theta_b = \sqrt{2}{\rm erf}^{-1}(1-2\gamma)$.

\subsection{Smoothness constraints}
Two additional constraints are defined to limit the variations of $P^m$ and $m_{k}^s$ between consecutive set-points time steps, for $k=0:N-1$,
\begin{align}
& |P_{k+1}^m -P_{k}^m| \leq \Delta P^m_{\max},  \label{eq:smooth_conmstr1} \\ 
& |m_{k+1}^s -m_{k}^s| \leq \Delta m^s_{\max}. \label{eq:smooth_conmstr2}
\end{align}

\subsection{The \ac{dap} algorithm}
Given a desired maximal failure rate ${\lambda}_{\max}$, the \ac{dap} algorithm consists in the solution of the following linear optimization problem:
\begin{equation}
\begin{aligned} 
J^* = &  \max_{\{P^m_k\}, \ \alpha, \  S_d^{\rm min}, \ S_d^{\rm max}} \sum_{k=0}^{N-1} c^{e}_k \tau P^m_k + c^{f} \alpha  \\   
& \mbox{subject to \eqref{eq:alphamin}, \eqref{eq:Es}--\eqref{eq:cSminmax}, \eqref{eq:SOEpv0}, \eqref{eq:SOEpvGauss2}--\eqref{eq:SOEpvGauss3}, \eqref{eq:msconstr1}--\eqref{eq:msconstr2},} \\
& \qquad \qquad \ \mbox{\eqref{eq:csoef1}--\eqref{eq:calphad1}, \eqref{eq:positivePm}, \eqref{eq:p_constraint1a}--\eqref{eq:p_constraint2a}, \eqref{eq:smooth_conmstr1}--\eqref{eq:smooth_conmstr2} } \nonumber
\end{aligned}
\end{equation}
The result of the optimization are the optimal \ac{is} base power profile $\{P^{md}_k\}=\{P^{m*}_k\}$ and the droop coefficient $\alpha^d=\alpha^*$, both defined the day before the delivery. The value of the cost function $J^{*}$ is equal to the day-ahead economical gain. 

\begin{figure}[t]
	\centering
	\includegraphics[width=\columnwidth]{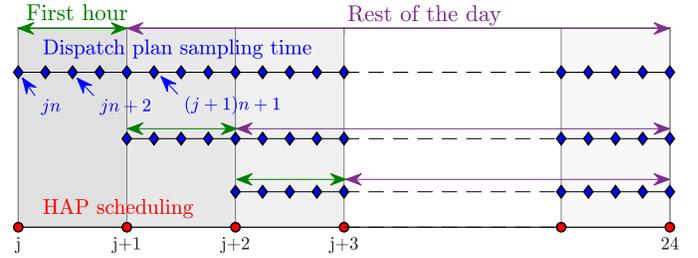}
	\caption{\ac{hap} time scheduling.}\label{fig:HAPtime}
\end{figure}

\section{Hours-Ahead Planning (HAP)}\label{sec:HAP}
The hour-ahead planning is a lower level controller which is re-computed every hour within the delivery day. The \ac{hap} routine receives from the \ac{dap} one the power delivery plan $\{P^{md}_k\}$ and the droop coefficient $\alpha^d$. The objective of \ac{hap} is to correct the plan $\{P^{md}_k\}$ to guarantee the provision of \ac{pfr}, keeping the droop coefficient $\alpha^d$ and reducing the expected \ac{dap} failure rate $\lambda_{\max}$ to a lower value $\lambda'_{\max}$, always maximizing the economical income.

Figure~\ref{fig:HAPtime} shows the \ac{hap} time scheduling. 
Let $j=0,1,\ldots,23$ indicate the hours during the day, and $n=3600/\tau$ be the number of intra-hour power set-points defined according to the dispatch plan sampling time.
Moreover, let $N_j = N-j\cdot n$ be the number of power set-points remaining from the $j$\textit{-th} hour to the end of the day.

At the beginning of  hour $j$, the \ac{is} power profile $\{P^{m}_k\}$ with $k=j n:N-1$ is re-programmed. Then, only the first $n$ steps, corresponding to the first hour of the dispatch plan, are applied. At hour $j+1$, the \ac{hap} optimization is repeated.
This time scheduling can be called \textit{reducing horizon}, and, similarly to the \textit{receding horizon} principle adopted by \ac{mpc}, it allows the control algorithm to be more robust with respect to modelling errors. In particular, at each hour, updated, and thus more accurate, \ac{pv} generation and \ac{pfr} energy requirement forecasts may be available, as well as the current value of the battery \ac{soe}. These updated data are useful to suitably correct the \ac{dap} program.

Based on this idea, as shown in Fig. \ref{fig:HAPtime}, the time from hour $j$ to the end of the day, is divided into two phases: the \ac{fh} ($k=jn:(j+1)n$), and the remaining time from hour $j+1$ to the end of the day ($k=(j+1)n:N$), from now named \ac{rod}.

At hour $j$, the available data are:
\begin{itemize}
\item[a)] the \ac{dap} power profile $\{ P^{md}_k\}$, $k=j\cdot n:N-1$ traded at the energy market;
\item[b)] the droop coefficient $\alpha^d$, defined for a given failure rate $\lambda^d$, to be guaranteed during all the day;
\item[c)] the updated \ac{pv}  forecasts $\{\hat{P}^{pv}_k\}$, with the associated standard deviations $\sigma^{pv}_k$, $k=j\cdot n:N-1$ (using the same the Gaussian model \eqref{eq:Pvgauss} adopted for \ac{dap});
\item[e)] the prediction of the frequency integral for the first hour $\widehat{W}^f_{h}$ and the associated standard deviation $\sigma^w_{1}$, computed as described in Section~\ref{sec:FrequencyForecast} with $T=$\SI{1}{\hour};
\item[f)] the prediction of the frequency integral for the rest of the day $\widehat{W}^f_{r}$ and the associated standard deviation $\sigma^w_{23-j}$, computed as described in Section~\ref{sec:FrequencyForecast} with $T=23-j$\si{\hour};
\item[g)] the penalty cost profile $\{ c^p_k \}$, $k=j\cdot n:N-1$ to be paid for a difference of the energy effectively exported by the \ac{is} from the energy traded at the day-ahead market;
\item[h)] the intra-day energy price profile $\{c^{i}_k\}$, $k=j\cdot n:N-1$;
\item[i)] the current battery \ac{soe}, $S_{jn}$. 
\end{itemize}

For both the time windows \ac{fh} and \ac{rod}, an approach similar to \ac{dap} is adopted. In particular, the basic idea of the partition of the \ac{bess} capacity by the definition of the thresholds $S^{\rm max}_d$ and $S^{\rm min}_d$ is re-applied with the definition of different thresholds: $S^{\rm max}_h$, $S^{\rm min}_h$, for the \ac{fh}, and $S^{\rm max}_r$, $S^{\rm min}_r$, for the \ac{rod}. The partition into two time windows is adopted in order to give more degrees of freedom to the optimization for the FH. Thanks to the use of short-term, and thus more accurate, predictions, the optimization over the FH will be finer. It is worth remarking that, as mentioned before, at each hour, the optimization results are applied only for the FH.

The \ac{hap} optimization problem, solved at each hour $j$, is formulated as it follows.
\begin{align}
&\max_{\{P^m_k\},   \mu_{h},  \mu_{r} }  L_j \nonumber\\
&L_j =  \sum_{k=jn}^{N-1}  (c^{i}_k-c^{p}_k) \tau (P^m_k - P^{md}_k)\delta^+_k  - c^{p}_k \tau  (P^m_k-P^{md}_k) \delta^-_k  \nonumber \\
                & \quad + w_h \mu_h + w_r \mu_r \label{eq:HAPcost}
\end{align}
\noindent{subject to:}
\begin{align}
& S^{\min} \leq S_{h}^{\min} \leq S_{h}^{\max}\leq S^{\max}, \label{eq:constr_1h}   \\
& m^s_{k} + \theta_{h} \sigma^{s}_{k} \leq S^{\max}_h    \qquad  \text{for } k = jn:j(n+1), \label{eq:constr_4h}  \\
- & m^s_{k} +\theta_{h} \sigma^{s}_{k} \leq S^{\min}_h   \qquad  \text{for } k = jn:j(n+1), \label{eq:constr_5h} \\
 &  2\alpha^d \widehat{W}_h^f = E_n[(S^{\rm max}+S^{\rm min})-(S^{\rm max}_h+S^{\rm min}_h)], \label{eq:constr_6h} \\
 &  2\alpha^d  \mu_h \sigma_{24}^w = E_{\rm n} [(S^{\max}-S^{\min})-(S^{\max}_h-S^{\min}_h)], \label{eq:constr_7h} \\
 & \mu \leq  \mu_h \leq \mu_{\max}, \label{eq:constr_8h} \\
& S^{\min} \leq S_{r}^{\min} \leq S_{r}^{\max}\leq S^{\max}, \\
& m^s_{k} + \theta_{r} \sigma^{s}_{k}  \leq S^{\max}_r  \qquad  \text{for }  k = j(n+1):N, \label{eq:constr_4r} \\
-& m^s_{k} +\theta_{r} \sigma^{s}_{k} \leq S^{\min}_r   \qquad  \text{for } k = j(n+1):N, \label{eq:constr_5r} \\
 &  2\alpha^d \widehat{W}_r^f = E_n[(S^{\rm max}+S^{\rm min}) -(S^{\rm max}_r+S^{\rm min}_r)], \label{eq:constr_6r} \\
 &  2\alpha^d \mu_r \sigma_{24}^w = E_{\rm n} [(S^{\max}-S^{\min})-(S^{\max}_h-S^{\min}_r)], \label{eq:constr_7r} \\
& \mu\leq  \mu_r \leq \mu_{\max}, \label{eq:constr_8r} \\
& m^s_{k} = S_{jn}- \frac{\tau  }{3600 \cdot E_{\rm n}} \sum_{i=jn}^{k-1} (P^m_i-\widehat{P}^{pv}_i), \label{eq:SOEpvGauss1_HAP}\\
& (\sigma_{k}^s)^2 = \left(\frac{\tau }{3600 \cdot E_{\rm n}} \right)^2 \cdot \sum_{i=jn}^{k-1} (\sigma_i^{pv})^2, \label{eq:SOEpvGauss2_HAP} \\
& P^m_k-\hat{P}^{pv}_{k}  + \alpha^d \Delta f_{\rm max}+\theta_b \sigma^{pv}_{k}\leq P^b_{\rm n} \label{eq:powerconstr_hap1}\\
& P^m_k-\hat{P}^{pv}_k  - \alpha^d \Delta f_{\rm max}-\theta_b \sigma^{pv}_{k}\geq -P^b_{\rm n}, \label{eq:powerconstr_hap2} \\
&  0\leq P^m_k \leq P^t_{\rm n}, \label{eq:positivepmHAP} \\
& |P_{k+1}^m -P_{k}^m| \leq \Delta P^m_{\max},  \label{eq:smoothnessHAP1} \\ 
& |m_{k+1}^s -m_{k}^s| \leq \Delta m^s_{\max}. \label{eq:smoothnessHAP2}
\end{align}


The optimization problem results to be mixed-integer with linear constraints. Indeed, there are two binary variables: $\delta^+_k$ defined (through  additive linear constraints not reported for clarity of presentation) to be equal to 1 when $P^m_k\geq P^{md}_k$ and 0 otherwise, and $\delta^-_k = 1-\delta^+_k$. 

For each of the two time windows, starting from the definitions of the new thresholds $S^{\rm max}_h$ and $S^{\rm min}_h$, for the \ac{fh}, and $S^{\rm max}_r$ and $S^{\rm min}_r$ for the \ac{rod}, the \ac{soe} constraints defined for \ac{hap} are reformulated as in \eqref{eq:constr_1h}--\eqref{eq:SOEpvGauss2_HAP}.

Let us focus on constraints \eqref{eq:constr_6h}--\eqref{eq:constr_7h} and \eqref{eq:constr_6r}--\eqref{eq:constr_7r}. They are the reformulation of the \ac{dap} constraints \eqref{eq:csoef1}--\eqref{eq:calphad1}, for the FH and the RoD, respectively. In \ac{dap}, \eqref{eq:csoef1}--\eqref{eq:calphad1} have to be respected in order to assure the maximal failure rate $\lambda^f_{\max}$ due to \ac{pfr}, which is related to coefficient $\mu$ by the relation $\mu~=~\sqrt{2}{\rm erf}^{-1}(1-\lambda^f_{\rm max})$ (see Section \ref{sec:FrequencyForecast}). It can be easily shown that $\mu$ increases when $\lambda^f_{\rm max}$ decreases. Therefore, if \eqref{eq:csoef1}--\eqref{eq:calphad1} are satisfied with a $\bar{\mu}\geq \mu$, the maximal failure rate $\lambda_{\max}$ is reduced. Indeed, by \eqref{eq:failure2} $\lambda_{\max}$ results to be reduced if $\lambda^f_{\max}$ decreases.
Constraints \eqref{eq:constr_6h}--\eqref{eq:constr_7h} for the FH and \eqref{eq:constr_6r}--\eqref{eq:constr_7r} for the RoD, are therefore re-formulated using the relevant predictions $\widehat{W}^f_{h}$ and $\widehat{W}^f_{r}$ and imposing that the droop coefficient $\alpha$ is equal to $\alpha^d$, computed by the \ac{dap}. 

Two optimization variables ${\mu}_{h}$ and ${\mu}_{r}$, are introduced for the \ac{fh} and \ac{rod} time-windows. The cost function \eqref{eq:HAPcost} is designed in order to increase their values, in order to obtain the reduction of the failure rate. With constraints \eqref{eq:constr_8h} and \eqref{eq:constr_8r}, ${\mu}_{h}$ and ${\mu}_{r}$ are limited by the minimal value $\mu$, which gives the guaranty to obtain the \ac{dap} failure rate $\lambda^f_{\rm max}$, and by the maximal value $\mu_{max}=\sqrt{2}{\rm erf}^{-1}(1-\bar{\lambda}^{f}_{\rm max})$, corresponding to the maximal reduced failure rate $\bar{\lambda}^{f}_{\max}<\lambda^f_{\max}$. 
The power and smoothness constraints \eqref{eq:powerconstr_hap1}--\eqref{eq:smoothnessHAP2} are re-written, as in \ac{dap}, for the entire interval $k=jn:N-1$, with $\alpha=\alpha^d$.

The cost function \eqref{eq:HAPcost} considers both the economical gain, determined by the balance between penalties and intra-day energy prices, and the reduction of the \ac{dap} failure rate, which, as mentioned, corresponds to the maximization of the coefficients $\mu_h$ and $\mu_r$. 
The optimization weights $w_h$ and $w_r$ have a different unit from the costs $c^{p}$ and $c^{e}$. Therefore, they has to be suitably normalized. It is worth remarking that the minimization of the failure rate may be in contrast with the maximization of the economical income. Therefore, the sizing of the weights $w_h$ and $w_r$ defines the priority level between the quality of the \ac{pfr} service and the economical gain.

\section{Simulation results} \label{sec:simulations}
A set of simulations has been performed considering real markets' data. 
The Italian day-ahead (MGP) and intra-day market (MI2) results (February 2019) \cite{MercatoElettricoITA} has been selected as input of \ac{dap} and \ac{hap} problems, respectively. The penalty for the variations on the dispatched power is fixed to \SI{0.05}{\EUR\per\kilo\Wh}. Moreover, the frequency regulating capacity price has been selected from the the results of the International \ac{pcr} markets between August 2018 and March 2019 \cite{internationalPCR}. 

\ac{dap} and \ac{hap} algorithms have been implemented in MATLAB/Simulink, and optimization problems have been written using the General Algebraic Modelling System (GAMS) language and solved with CPLEX.
Battery is modelled with a standard equivalent circuit in which the internal resistance is a function of the \ac{soe} and of the electromotive force. Thus, a variable nonunitary battery efficiency has been implemented. 

Inputs of the simulator are real \ac{pv} measurements and \ac{pv} forecasts registered is the low-voltage (LV) microgrid realized by the University of Genova \cite{adinolfi2015advanced}.
Moreover frequency measurements from the UK grid has been adopted in the construction of the \ac{ar} models and for the simulations \cite{DatiFreq}. 

Simulations have been executed over a 21 days period and considering the implementation only of \ac{dap}, and of both \ac{dap} and \ac{hap}.
Moreover, five different cases are proposed, characterized by different \ac{pv}-\ac{bess} sizes, as reported in Table~\ref{tab:simresults}. Considering devices rating, the \ac{isms} is expected to differently balance the two services, \textit{i.e.} a larger \ac{bess} will provide higher regulating capacity but can rely on smaller offsets for charge management, on the other hand, a larger \ac{pv} will drive the \ac{isms} to privilege the dispatch service.

Table~\ref{tab:simparam} shows the parameters adopted for the \ac{is}. Among the others: the minimum droop coefficient $\alpha_{\min}$ is defined according to \eqref{eq:alphamin} with respect to the \ac{pv} nominal power, with an equivalent maximal statism $b_p^{\max}$ fixed to \SI{8}{\percent} \cite{aisbl2012entso}; the maximum failure rate $\lambda_{\rm max}$ is fixed at \SI{5}{\percent}, according for example to the requirements of the UK market \cite{DatiFreq,EFR_Uk}; the dispatch sampling time $\tau$ is set to \SI{15}{\minute} according to the Italian energy market\cite{MercatoElettricoITA}.

\begin{figure}[t]
	\centering
	\includegraphics[width=\columnwidth]{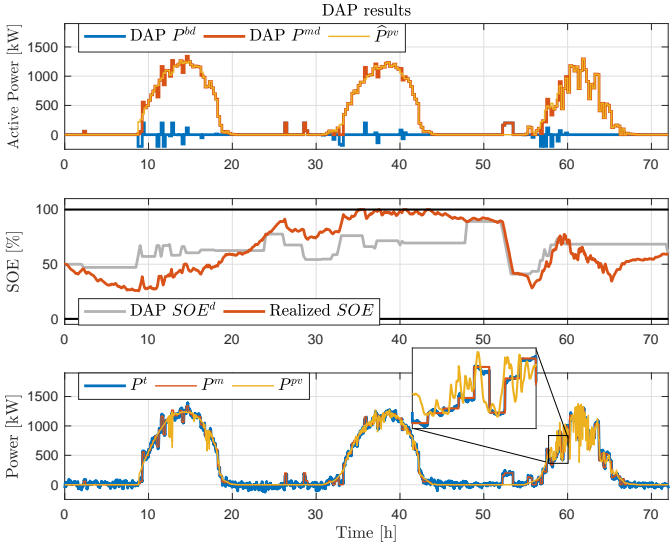}
	\caption{Simulation results for the \ac{dap} only configuration, Case A: \SI{1500}{\kilo\watt} \ac{pv}, \SI{500}{\kilo\Wh} \ac{bess}. Top: planning power profiles; middle: planned and realized \ac{soe} profiles; bottom: realized power profiles.}\label{fig:DAP}
\end{figure}

\begin{table}[t]
		\centering
		\caption{Simulation results.}
		\label{tab:simresults}
		 \renewcommand{\arraystretch}{1.4}
        	
        \begin{tabular}{c c c c c c  }
        \hline \hline
            Case   & $\lambda$ \si{\percent}  & Total \euro{} & \ac{pcr} \euro{}   & Dispatch \euro{} & Penalty \euro{} \\ \hline
            \multirow{2}{*}{A}  & 0.424 & 27798  & 15756 & 12042  & 0 \\ 
                                & 0     & 24825  & 14483 & 11363  & -1020 \\ \hline
            \multirow{2}{*}{B}  & 0.530 & 35949  & 23940 & 12010  & 0 \\ 
                                & 0     & 32150  & 22575 & 11059  & -1484 \\ \hline
            \multirow{2}{*}{C}  & 0.403 & 38765  & 14420 & 24345  & 0 \\ 
                                & 0     & 36997  & 14035 & 23797  & -835 \\ \hline
            \multirow{2}{*}{D}  & 0.234 & 41536  & 4821  & 36714  & 0 \\  
                                & 0     & 40897  & 4957  & 36426  & -486 \\ \hline
            \multirow{2}{*}{E}  & 0.941 & 41104  & 4274  & 36830  & 0 \\
                                & 0     & 40451  & 4287  & 36580  & -416 \\\hline
            \multicolumn{6}{c}{
            \begin{tabular}{p{0.4\textwidth}}
            Resources sizes: A. \ac{pv} \SI{500}{\kilo\watt}, \ac{bess} \SI{1500}{\kilo\watt}; B. \ac{pv} \SI{500}{\kilo\watt}, \ac{bess} \SI{1000}{\kilo\watt};  C. \ac{pv} \SI{1000}{\kilo\watt}, \ac{bess} \SI{1000}{\kilo\watt}; D. \ac{pv} \SI{1500}{\kilo\watt}, \ac{bess} \SI{500}{\kilo\watt}; E. \ac{pv} \SI{1500}{\kilo\watt}, \ac{bess} \SI{320}{\kilo\watt}.
            \end{tabular}
            }      
            \\ \hline \hline
        \end{tabular}
\end{table}

\begin{figure}
	\centering
	\includegraphics[width=\columnwidth]{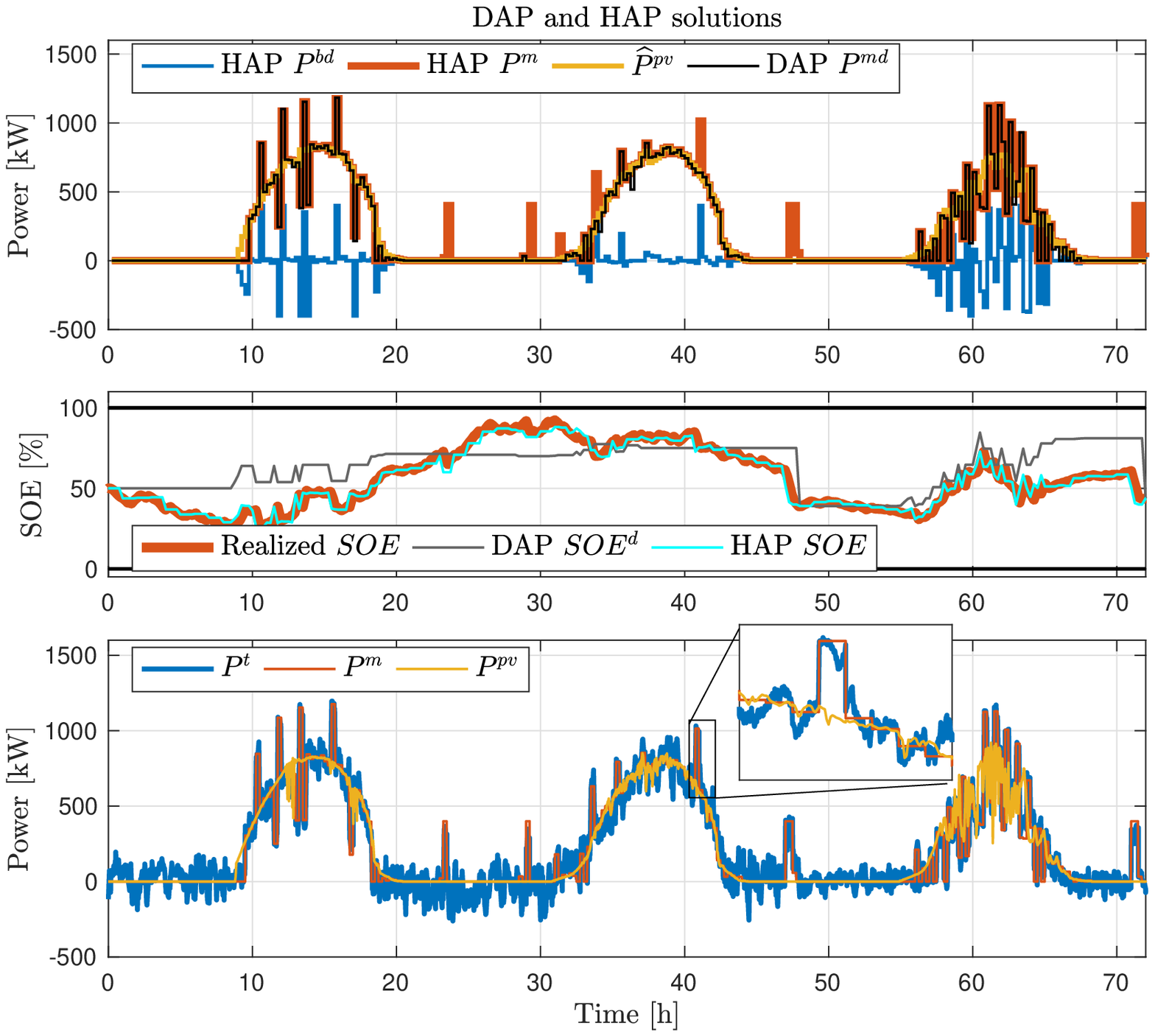}
	\caption{Simulation results for the \ac{dap}-\ac{hap} configuration, Case C: \SI{1000}{\kilo\watt} \ac{pv}, \SI{1000}{\kilo\Wh} \ac{bess}. Top: planning power profiles; middle: planned and realized \ac{soe} profiles; bottom: realized power profiles.}\label{fig:DAPeHAP}
\end{figure}

\begin{table}
		\centering
		\caption{Simulation parameters.}
		\label{tab:simparam}
		 \renewcommand{\arraystretch}{1.4}
		 \begin{tabular}{l l l}
		 \hline \hline
		 Variable                 & Description                               & Value \\\hline 
		 $\tau$                   & Dispatch sampling time                    & \SI{15}{\minute}\\
		 $\Delta P^m_{\max}$      & Maximal power deviation                   &  \SI{40}{\percent}$P^t_{\rm n}$\\
		 $\Delta m^s_{\max}$      & Maximal \ac{soe} deviation                & \SI{10}{\percent}\\
		 $\gamma$                 & Battery power chance-contraints coefficient          & \SI{1}{\percent}\\
		 $\beta$                  & Battery \ac{soe} chance-contraints coefficient      & \SI{1}{\percent}\\
		 $\alpha_{\min}$          & Minimal droop coefficient as \eqref{eq:statism_max} with $b_p^{\rm max}$ \SI{8}{\percent} &  - \\
		 $\alpha_{\max}$          & Maximal droop coefficient   & inf \\
		 $S^{\min}$               & Maximal battery \ac{soe}  & \SI{100}{\percent}\\
		 $S^{\max}$               & Maximal battery \ac{soe}   & \SI{0}{\percent}\\
		 $\Delta f^{\max}$        & Maximal frequency deviation  & \SI{0.2}{\hertz}\\	 
		 $\mu$                    & Equivalent to \ac{dap} failure rate $\lambda_{\rm max}=$\SI{5}{\percent}  & 1.96 \\	 
		 $\bar{\mu}_{\max}$       & Equivalent to \ac{hap} failure rate $\bar{\lambda}_{\rm max}=$\SI{0.3}{\percent}  & 3  \\	
		 \hline \hline
        \end{tabular}
\end{table}

Figure~\ref{fig:DAP} shows a section of the simulation of the stand alone \ac{dap} controller. The top plot reports the dispatch plan $\{P^{md}_k\}$, the day ahead \ac{pv} forecast $\{\hat{P}^{pv}_k\}$ and the battery offset program $\{P^{bd}_k\}=\{P^{md}_k-\hat{P}^{pv}_k\}$. The middle plot depicts the programmed \ac{soe} trajectory and the realized ones. While the bottom plot shows the resulting profiles of the total power at the \ac{gcp} $P^t$, of the base dispatch power $P^m$ and of the \ac{pv} generation $P^{pv}$. 

The detailed numerical results of all the simulations in the the stand alone \ac{dap} case are reported in Table~\ref{tab:simresults}. The reported data show that the \ac{dap} is able to determine a reliable power profile, which allows the \ac{is} to perform both the services with a failure rate lower than the prescribed maximal value $\lambda_{\rm max}=5\%$.

\begin{figure}
	\centering
	\includegraphics[width=\columnwidth]{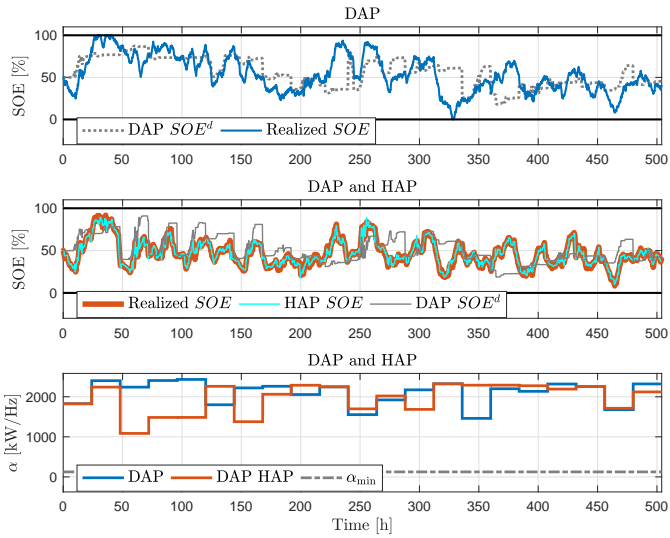}
	\caption{Simulation results for the \ac{dap} only and \ac{dap}-\ac{hap} configurations, Case D: \SI{500}{\kilo\watt} \ac{pv}, \SI{1500}{\kilo\Wh} \ac{bess}. Top:  planned and realized \ac{soe} profiles in the stand alone \ac{dap} configuration; middle: planned and realized \ac{soe} profiles in the \ac{dap}-\ac{hap} configuration; bottom: droop coefficients obtained in the \ac{dap} only and \ac{dap}-\ac{hap} configurations.} \label{fig:DAPHAPconfr}
\end{figure}

Figure~\ref{fig:DAPeHAP} shows an example of the results obtained with the \ac{dap}-\ac{hap} configuration. In particular, in the top plot the modification operated by \ac{hap} with respect to \ac{dap} can be appreciated. For example, during the night operations (from hour 20 to hour 31) the \ac{hap} commands some short power delivery in order to discharge the battery and avoid to reach the full charge condition. Also Fig.~\ref{fig:DAPHAPconfr} makes evidence on the advantages on using the \ac{hap} procedure. Indeed, with the stand alone \ac{dap}, during the first 50 hours, the battery \ac{soe} reaches the up limit (failure), whereas this does not happen when \ac{hap} is used. It can be observed that in all the considered cases the \ac{dap}-\ac{hap} strategy allows obtaining a null failure rate, as shown in Table~\ref{tab:simresults}. It is worth remarking that one of objective of the \ac{hap} is to reduce the expect failure rate to a value below \SI{1}{\percent}.

The bottom plot of Fig.~\ref{fig:DAPHAPconfr} reports the droop coefficients computed with the two configurations. They result to be comparable, even if the \ac{dap} solution allows to reach slightly higher values. As a consequence, the total economical income results to be higher. It is worth remarking that this results are not affected by some penalty that could be payed for reaching fail conditions in the \ac{hap} case.

The results reported in Table~\ref{tab:simresults} prove the effectiveness of the control algorithms with all the different considered configurations. All cases use the same price vectors, therefore, the power ratings of the \ac{is} has a
relevance on the total income. Increasing the \ac{pv} power rating allows to reach higher income from the dispatch, while the highest regulating capacities are obtained with larger \acp{bess}.

It is finally worth remarking that, as noticed in Section~\ref{sec:ProblemDefinitions}, the control algorithms consider a battery with unitary efficiency. On the contrary, the test battery model adopted for the tests account for the efficiency. Such a model has been derived from the simulation setup presented and validated in \cite{PSCC2018,powertech2019}. The model consists in the series of an internal voltage source and of variable resistance, the parameters obtained from measurements the original grid-scale lithium-titanate battery rated \SI{560}{\kWh} \cite{sossan2016Achieving} has been scaled to match the different battery sizes simulated. The obtained results prove that such an approximation in the control design does not influence the overall performance. 

\section{Conclusions} \label{sec:conclusions}
This paper presents a strategy for the optimal planning of an integrated \ac{bess}--\ac{pv} system, which provides frequency regulation and generation dispatch. The control architecture is composed by two algorithms. The first one, \ac{dap}, is executed the day before the delivery and defines the power dispatch plan and a droop coefficient for the \ac{pfr}, on the basis of \ac{pv} forecasts and predictions of the energy required for providing \ac{pfr}. The delivery day, at each hour, the second algorithm, named \ac{hap}, is executed in order to allow the \ac{is} to perform its tasks in a continuous and reliable way by using updated short-term forecasts. The two algorithms are designed to maximize the total incomes and the performance in providing \ac{pfr}. They use chance-constrained optimization in order to model the forecasts errors. The control framework has been validated by simulations.
Future works will consider different applications using a similar approach, also non-Gaussian representations of uncertainties and stochastic models of the energy prices.

\appendices
\section{Proof of Proposition \ref{pr:1}}
Using \eqref{eq:SOEdec1}, from \eqref{eq:chanceSOEpv} and \eqref{eq:chanceSOEf}, it follows that
$$
\mathbf{P}\left(0\leq S^{pv}_k-S^{\min}_d \leq \frac{E_{\rm n}^{pv}}{E_{\rm n}}\right) = \mathbf{P}(A) \geq 1-\beta,
$$
$$
\mathbf{P}\left(0\leq S^f_k-S^{\min} \leq \frac{E_{\rm n}^{f}}{E_{\rm n}}\right)= \mathbf{P}(B)  = 1-\lambda^f_{\max}, \nonumber
$$
where, $A$ and $B$ indicate the two considered constraints. Since ${S}^{pv}_k$ and ${S}^{f}_k$ are independent, it results that 
$$
\mathbf{P}(A \cap B) =  \mathbf{P}(A)\cdot \mathbf{P}(B) = (1-\beta)\cdot (1-\lambda^f_{\max}) = 1-{\lambda}_{\max}  
$$
where ${\lambda}_{\max}$ is equal to the one defined in \eqref{eq:failure2}.
Now consider that, because of elementary set inclusion properties,
\begin{equation}
\begin{aligned}
&\mathbf{P}\left(0\leq S^{pv}_k+S^f_k-S^{\min}_d-S^{\max} \leq \frac{E_{\rm n}^{pv}}{E_{\rm n}} + \frac{E_{\rm n}^{f}}{E_{\rm n}} \right) \nonumber \\
 & \qquad \qquad \qquad\qquad \qquad\qquad \geq \mathbf{P}(A \cap B) \geq 1-{\lambda}_{\max}  
\end{aligned}
\end{equation}
from which, taking into account \eqref{eq:SOEdec}, it follows that
$$
\mathbf{P}\left(0\leq S_k - S^{\rm min} \leq \frac{E_{\rm n}^{pv}+E_{\rm n}^{f}}{E_{\rm n}}\right) \geq 1-{\lambda}_{\max}  
$$
and, therefore,
$$
\mathbf{P}\left(S^{\rm min}\leq S_k  \leq \frac{E_{\rm n}^{pv}+E_{\rm n}^{f}}{E_{\rm n}}+ S^{\rm min} \right) \geq 1-{\lambda}_{\max}.  
$$
To conclude, \eqref{eq:failure1} is proved by noticing that from the definitions \eqref{eq:Epvn} and \eqref{eq:Efn} it results that
$$
\frac{E_{\rm n}^{pv}+E_{\rm n}^{f}}{E_{\rm n}}+ S^{\rm min} = \frac{E_{\rm n}^{s}}{E_{\rm n}}+ S^{\rm min} =  S^{\rm max}.
$$

\ifCLASSOPTIONcaptionsoff
  \newpage
\fi

\end{document}